# Filament based Ionizing Radiation Sensing Technology

Weiwei Liu[1,2], Jiewei Guo[1,2], Nan Zhang[1,2], Lu Sun[1,2], Haiyi Liu[1,3], Shihi Tao[1,2], Yuezheng Wang[1,2], Binpeng Shang[1,3], Pengfei Qi[1,2]\*, Lie Lin[1,3]

[1] *Institute of Modern Optics, Eye Institute, Nankai University, Tianjin 300350, China*

[2] *Tianjin Key Laboratory of Micro-scale Optical Information Science and Technology, Tianjin 300350,China*

[3] *Tianjin Key Laboratory of Optoelectronic Sensor and Sensing Network Technology, Tianjin 300350, China*

\* [qipengfei@nankai.edu.cn](qipengfei@nankai.edu.cn)

**Abstract:** Accidental exposure to overdose ionizing radiation will inevitably lead to severe biological damage, thus detecting and localizing radiation is essential. Traditional measurement techniques are generally restricted to the limited detection range of few centimeters, posing a great risk to operators. The potential in remote sensing makes femtosecond laser filament technology great candidates for constructively address this challenge. Here we propose a novel filament-based ionizing radiation sensing technology (FIRST), and clarify the interaction mechanism between filaments and ionizing radiation. Specifically, it is demonstrated that the energetic electrons and ions produced by $\alpha$ radiation in air can be effectively accelerated within the filament, serving as seed electrons, which will markedly enhance nitrogen fluorescence. The extended nitrogen fluorescence lifetime of ~1 ns is also observed. These findings provide insights into the intricate interaction among ultra-strong light filed, plasma and energetic particle beam, and pave the way for the remote sensing of ionizing radiation.

**INTRODUCTION**

Ionizing radiation (IR), energy flux in the form of atomic and subatomic particles or electromagnetic waves, is a perennial hot topic in a variety of fields, such as atomic physics, nuclear safety, and radioactive medicine.  A long list of historic experiments and instruments demonstrates its special importance in the understanding of fundamental phenomena and have been milestones in the building up of modern theories[1]. However, accidental exposure to overdose ionizing radiation will inevitably lead to severe biological damage, thus detecting and localizing radiation is the very basis of physicists work. Traditional measurement techniques are restricted to the limited detection range of few centimeters[2], posing a great risk to operators. Therefore, developing non-contact IR detection technology is imperative.

Recently, the rapidly developing femtosecond laser filament technology provides a new opportunity for the remote sensing of IR. Due to the dynamic balance between the optical Kerr self-focusing effect and the plasma defocusing effect, the propagation of high-power femtosecond lasers in atmosphere can induce a unique nonlinear optical phenomenon, filamentation [3-6]. During filamentation in the atmosphere, the ultra-strong field of $10^{13}$–$10^{14}$ W/cm$^2$ with a large distance ranging from meter to kilometers can effectively ionize, break, and excite the molecules and fragments, resulting in characteristic fingerprint emissions, which provide a great opportunity for investigating strong-field molecules interaction in complicated environments, especially remote sensing [7-9]. The length, localization, and intensity of the femtosecond laser filament can be effectively controlled by initial laser parameters and the spatiotemporal shaping [10-14]. Therefore, the versatile femtosecond laser filament technology demonstrates a great potential for the remote sensing of air pollutants and hazardous goods such as explosives and chemical weapons. In principle, IR can effectively disrupt molecular bonds and generate the energetic electrons and ions, significantly affects the femtosecond laser filamentation, illustrating the feasibility of the remote sensing of IR by filaments. However, such critical detecting strategy haven't yet been reported. The physical mechanism of the interaction between femtosecond laser filamentation and IR is not clear.

As a typical IR, helium nuclei ($\alpha$ particle) emitted by some radioactive materials gain considerable fame since they were identified and applied to explore atomic structure by Ernest Rutherford[15]. Meanwhile, $\alpha$ particles have been widely used in various fields, such as nuclear physics[16], astrophysics[17,18], and medical science[19,20], due to their unique properties of high ionization density and short penetration depth. Here we shed light on the interaction between femtosecond laser filaments and $\alpha$ particles in experimental and theoretical, as well as propose the filament based ionizing radiation sensing technology (FIRST). It is demonstrated that the substantial energetic electrons and ions generated from α radiation in air can be accelerated within the filament as seed electrons, leading to an increase in the intensity of backward nitrogen fluorescence signals. Hence, it is promising to achieve remote sensing of α radiation, laying a firm foundation for detecting IR in civilian and military nuclear facilities or equipment.

## RESULTS AND DISCUSSION

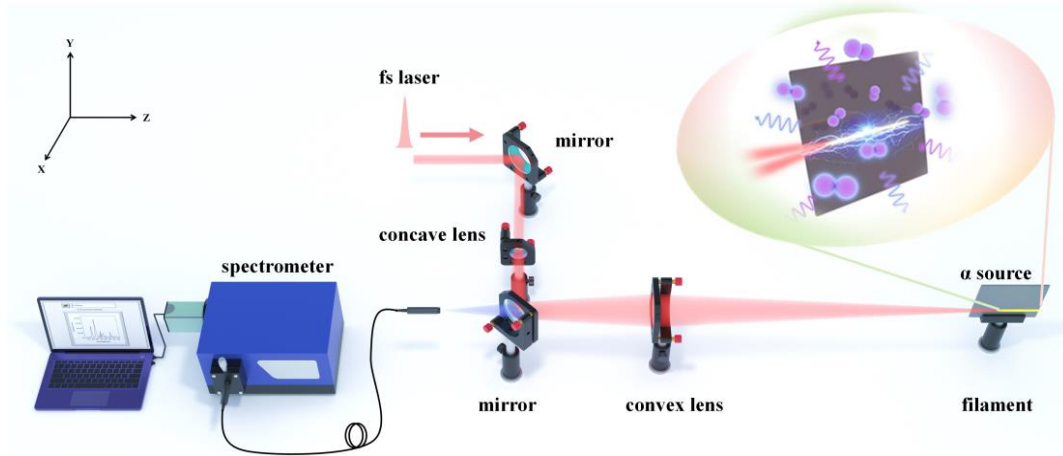

**Fig. 1. Schematic diagram for FIRST**.

Fig. 1 illustrates the schematic diagram for FIRST. A commercial femtosecond laser system (Legend Elite, Coherent Inc.) delivers 60 fs pulses at central wavelength of 800 nm with maximum pulse energy of 3.5 mJ at a repetition rate of 500 Hz. The laser pulse output from the laser system was focused by a lens group comprising of a concave lens with a focal length of -200 mm and a convex lens with a focal length of 500 mm. The geometrical focus of the lens group was located 1 m away from the convex lens. The resulting femtosecond laser filament was initiated around the geometrical focus of the lens group and extended over a length of more than 150 mm. The safe-grade α-plate radiation source (241Am, half-life of 432a, size 150 mm×100 mm) is placed vertically on one side of the filament, and the distance between the α-plate radiation source and the light filament can be adjusted by a precise electric control stage. The backward fluorescence emitted by the plasma filament is collected by the f = 500 mm lens into the fiber tip of a spectrometer (Omni-$\lambda$ 300，Zolix Ltd.) equipped with an intensified CMOS camera (Istar-sCMOS, Andor Technology Ltd.).

A wide-angle lens equipped CCD camera was positioned perpendicular to the femtosecond laser filament to characterize its properties. The inset in Fig. 2a presents a lateral view of the filament in the experiment where the laser propagates from left to right. Fig. 2 depicts the normalized intensity distribution of filament along the propagation direction. The extent of filament can be defined by the intensity drops to $1/e^2$ of peak value. Clearly, the filament length is approximately 15 mm under the laser parameters and focusing conditions. Fig. 2b presents the recorded fluorescence spectrum of the filament without $α$ radiation source. The ultraviolet emission can be attributed to excited nitrogen molecules and ions, and the quantum vibrational numbers of the upper and lower levels of each optical transition are denoted by the numbers in parenthesis.

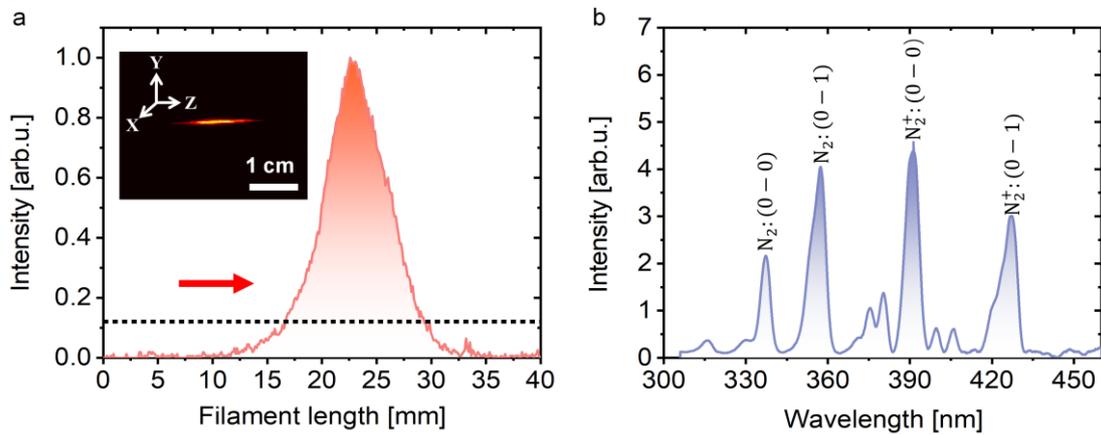

**Fig. 2. Femtosecond laser filamentation.** (a) Normalized fluorescence intensity of filament along propagation axis (the inset is side fluorescence image of filament); (b) Backward fluorescence spectrum of nitrogen induced by filament without $α$-plate radiation source.

Then the effect of IR on the fluorescence emission during filamentation was investigated. Fig. 3a depicts the recorded nitrogen fluorescence spectra for different distances between the $α$-plate radiation source and the filament, which can significantly impact the IR intensity within the filament, thereby modulate the filamentation dynamics and the fluorescence emission. It can be clearly observed that the increased distance between the $α$-plate radiation source and the filament reduce the intensity of fluorescence emission. To confirm the varying nitrogen fluorescence spectra stem from the IR effects of $α$ particles, a blank contrast plate (i.e., without $α$-particle radiation) with parameters identical to the aforementioned radiation source was adopted for comparative experiments under same experimental conditions, and the corresponding results are presented in Fig. 3c. The distance between the blank contrast plate and the filament does not lead to any appreciable changes in nitrogen fluorescence signal intensity. It was corroborated that the nitrogen fluorescence spectra modulated by the distance between the filament and the α-plate radiation source in Fig. 3a arises from the IR effects of α particles. For clarity, the integrated intensity of nitrogen fluorescence at 337 nm and 391 nm were extracted and plotted in Figs. 3c and 3d, corresponding to Figs. 3a and 3b, respectively. The red region in the figure clearly shows that the α-particle source can significantly enhance the nitrogen fluorescence intensity at the emission peaks of 337 nm and 391 nm, within a distance of 1.5 cm from the filament. Due to the short penetration length of α particles in air, the backward nitrogen fluorescence intensity does not change significantly for the distance greater than 1.5 cm.

Considering the distance between the target and the apparatus for femtosecond laser filamentation and fluorescence detection can reach kilometers in previous reports, the remote and non-destructive detection of α-particle sources is feasible based on femtosecond laser filamentation technology. Huggins et al. first proposed the α particle induced ultraviolet (UV) luminescence of air[21], suggesting the potential non-contact indirect optical sensing of alpha particles[22,23]. Because the induced air fluorescence is mainly emitted from the excited nitrogen ions and molecules by α radiation, with a emission band of 300~400 nm[24]. Therefore, current optical methods for the detection of α-particles are focused on these UV-radiations. It is worth noting that the solar radiation in this range can reach $10^{-12}$ W·cm$^{-2}$·nm$^{-1}$[25,26], so in the natural environment, the UV background intensity is usually greater than the IR-induced fluorescence, which masks the characteristic information and brings a significant interference to the sensitivity of this technique. Currently, most IR-based α-particle detection is generally limited to dark environments. For example, Inrig et al. achieved accurate detection of α radiation sources under very harsh conditions by irradiating a non-UV light source at 1.5 m from the detector for 10 s[22]. It was not until 2016 that Sand et al. demonstrated remote detection of α sources at a distance of 1 m and a measurement time of 10 s under normal illumination conditions[27]. In contrast, our results are instructive for detecting α radiation sources at greater distances and in faster times.

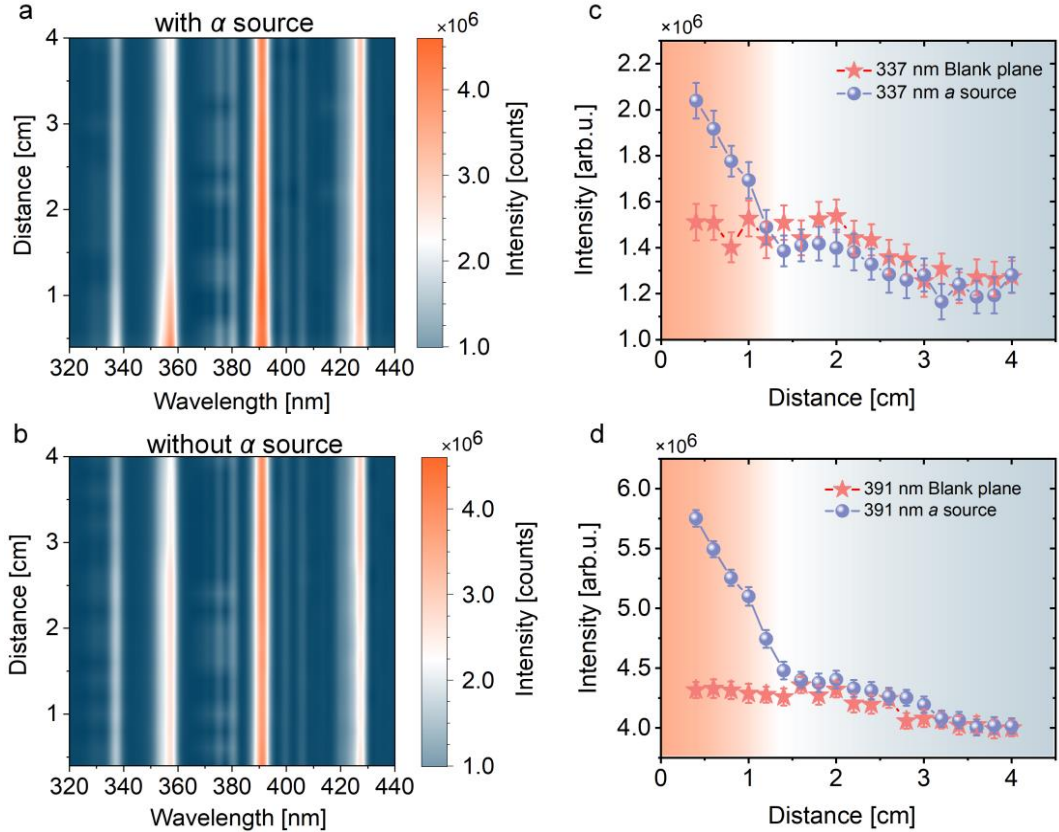

**Fig. 3. Enhanced nitrogen fluorescence emission by *α*-plate radiation source.** (a), (b) Nitrogen fluorescence spectra for different distances between filament and planes with/without *α* radiation source; (c), (d) Distance dependent total nitrogen fluorescence emission centered at (b)337 nm and (d) 391 nm.

To clarify the underlying mechanism of the enhanced fluorescence emission by IR during filamentation, the IR effect on the lifetime of nitrogen fluorescence emission is an important supplement. Consequently, then the time resolved fluorescence spectra for the two schemes in Figs. 3a and 3b were measured by the combination of gated iCMOS and spectrometer, as shown in Fig. 4. During data acquisition, the gate width of 1 ns with a step of 1 ns were adopted. It can be clearly observed that the nitrogen fluorescence lifetime can be extended by about 1 ns through the *α*-particle planar radiation source, which leading to an observable enhancement of fluorescence intensity in Fig. 3.

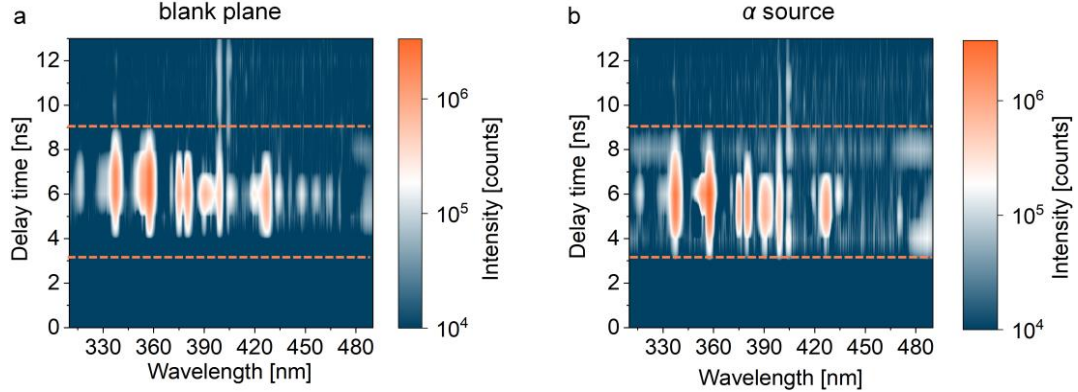

**Fig. 4. Extended nitrogen fluorescence lifetime by *α*-plate radiation source.** Time resolved fluorescence spectra for (a) blank contrast board and (b) *α* plane radiation source.

As shown in Fig. 2b, the emission of nitrogen fluorescence spectrum during femtosecond laser filamentation are mainly determined by the first negative band system ($B^2\Sigma_u^+ \to X^2\Sigma_g^+$ transition) of $N_2^+$ and the second positive band system ($C^3\Pi_u^+ \to B^3\Pi_g^+$ transition) of $N_2$ [28-32]. The emission line at 337 nm is generated by the excitation of $N_2$, while the emission line at 391 nm originates from $N_2^+$ [33]. As for $N_2^+(B^2\Sigma_u^+)$, it can be generated through two possible schemes: (1) photo-ionization (multiphoton ionization or tunnel ionization) of the inner electron of $N_2$ in the strong laser field; (2) ionization of $N_2$ via the collision with electron with high kinetic energy[34,35]. However, the predominant line of which at 337.1 nm is due to a transition between excited states $C^3\Pi_u^+$ and $B^3\Pi_g^+$ of the triplet manifold of neutral nitrogen molecules, in which the excitation of neutral $N_2$ occurs via multiple collisions with free electrons[36,37].

It was demonstrated that plenty of energetic electrons can be produced during alpha particles colliding with air molecules[38]. These kinetic electrons can continuously collide and excite atmospheric molecules, mainly nitrogen and oxygen molecules, and eventually generate fluorescence. Therefore, the physical scenario of the enhanced nitrogen fluorescence during femtosecond laser filamentation by alpha particle radiation can be summarized as shown in Fig. 5. In the presence of alpha radiation, the air molecules ionized and excited by alpha particles can provide a considerable amount of free electrons and nitrogen ions, which can serve as seed electrons to be accelerated by light field of laser pulse, enhance the impact ionization during filamentation. Then the kinetic electrons collide with the excited nitrogen molecular ions can produce $N_2^+(B^2\Sigma_u^+)$ and $N_2(C^3\Pi_u)$ through following process[37,39]:

$$N_2(X^1\Sigma_g^+) + e^- \rightarrow N_2(C^3\Pi_u) + e^- \tag{1a}$$

$$N_2(X^1\Sigma_g^+) + e^- \rightarrow N_2(B^2\Sigma_u^+) + 2e^- \tag{1b}$$

The increased population of $N_2(C^3\Pi_u)$ and $N_2^+(B^2\Sigma_u^+)$ ultimately lead to the enhanced nitrogen fluorescence emission.

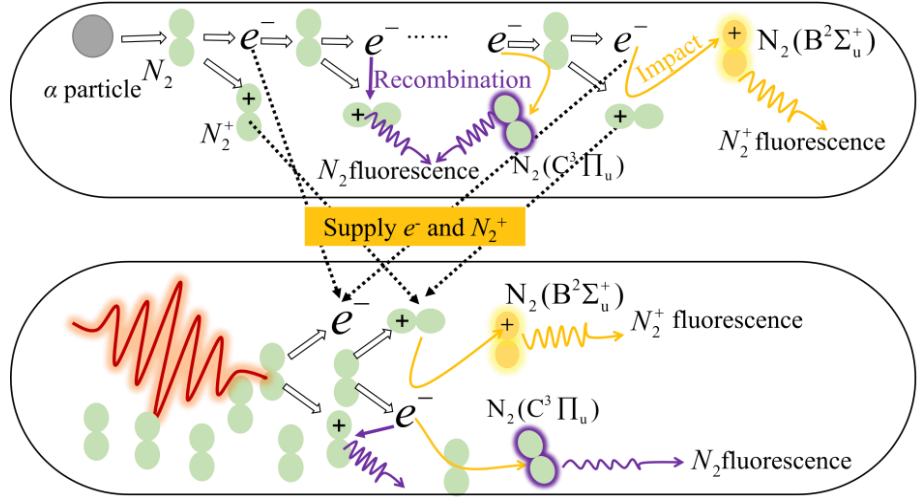

**Fig. 5 Physical scenario of the enhanced nitrogen fluorescence by *α* particles during femtosecond laser filamentation.**

Lastly, we further quantitatively verify the physical scenario of the enhanced nitrogen fluorescence by *α* particles through numerical simulation. The $N_2^+(B^2\Sigma_u^+)$ emitting ultraviolet photons of 391 nm could be generated from the path: ionization of $N_2$ via the collision with electron with high kinetic energy, i.e., Eq. 1[37]. While the $N_2(C^3\Pi_u)$ could be generated through the possible scheme: electron impact excitation of $N_2$ by electron with high kinetic energy, i.e., Eq. 2.[37] For a more quantitative analysis, we simulate the formation dynamics of excited neutral molecules using a rate equation for the neutral nitrogen excitation processes, taking $N_2(C^3\Pi_u)$ as an example:

$$\frac{dN_2(C^3\Pi_u)}{dt} = \langle\sigma\upsilon\rangle n_e N_2 - \frac{N_2(C^3\Pi_u)}{\tau_c} - k_1 N_2 N_2(C^3\Pi_u) \tag{2a}$$

$$\frac{dn_e}{dt} = \alpha n_e + k_2 N_4^+ n_e - k_3 N_2^+ n_e - \eta n_e \tag{2b}$$

$$\frac{dN_4^+}{dt} = k_3 N_2^+ N_2 - k_2 n_e N_4^+ \tag{2c}$$

$$\frac{dN_2^+}{dt} = \alpha n_e - k_3 N_2^+ N_2 - k_3 N_2^+ n_e \tag{2d}$$

In the above equations, cross section $\sigma$ is analytically derived from the reference [40] while assuming a Maxwellian distribution for electron velocity in order to compute $\langle \sigma v \rangle$ [41,42]. And the $\tau_c = 25$ ns is the radiative lifetime, $k_1 = 3 \times 10^{-10}$ $cm^3/s$ the quenching rate of the triplet state through collisions with molecular nitrogen[41], $\alpha = 7.6 \times 10^{11} \{[T_e(eV)]/[15.6]\}^{1.5} \times \{[15.6]/[T_e(eV)] + 2\} e^{-15.6/T_e(eV)} s^{-1}$ is the ionization rate, $k_2 = 2.6 \times 10^{-6} [T_e(K)/300]^{-0.5}$ $cm^3/s$ is the rate of electron capture by $N_4^+$, $k_3 = 4.3 \times 10^{-8} [T_e(eV)]^{-0.4}$ $cm^3/s$ is the rate of electron capture by $N_2^+$, $\eta = 2.75 \times 10^{-10} T_e^{-0.5} e^{-5/T_e} N_2 + 1.5 \times 10^{-32} T_e^{-0.5} e^{-0.052/T_e} N_2^2 s^{-1}$ is the attachment rate.

The next step is to combine the ionization effect of $\alpha$ particles with the many body evolution during filamentation described by Eq. (2). The ionization energy loss rate of $\alpha$ particles in air can be calculated by the following Bethe's formula:[43]

$$-\frac{dE}{dr} = \frac{4\pi z^2 e^4 ZN}{m_0 v^2} [\ln(\frac{2m_0 v^2}{I})](\frac{1}{4\pi\varepsilon_0})^2 \tag{3}$$

where $v$ is the particle flight velocity, Z is the atomic number of the medium, $ze$ is the charge of the charged particle, $m_0$ is the rest mass of the electron, N is the number of atoms per unit volume in the medium, and $I$ is the average ionization potential, which represents the average of the excitation and ionization energies of the electrons in each shell layer. We assume that the energy loss of an alpha particle per unit length along its propagation path is arising from the atmosphere ionization, accompanied by the generation of electrons and ions. The quantity of generated electrons can be given by $-dE/dr = \varepsilon n_0$, and the electron density along the propagation path can be calculated as: $n = n_0/r^2$, where r is length from alpha radiation source to filament. The $\alpha$ particles induced electrons at the location of the femtosecond laser filament are the integration along the entire propagation path, i.e., $N_e = \frac{1}{2\pi} \int_0^{r'} n e^{-u(r-r')} dr'$, where $r'$ is the length from $\alpha$ particles to filament. Additionally, since the $\alpha$ radiation source is a plane, it is necessary to integrate over the effective area of the radiation source plane, which can be described by $N_{eR} = \int_0^{R_{max}} N_e 2\pi R dR$, where $R_{max} = \sqrt{d^2 - r^2}$ denotes the radius of the effective area of action of the $\alpha$ radiation source.

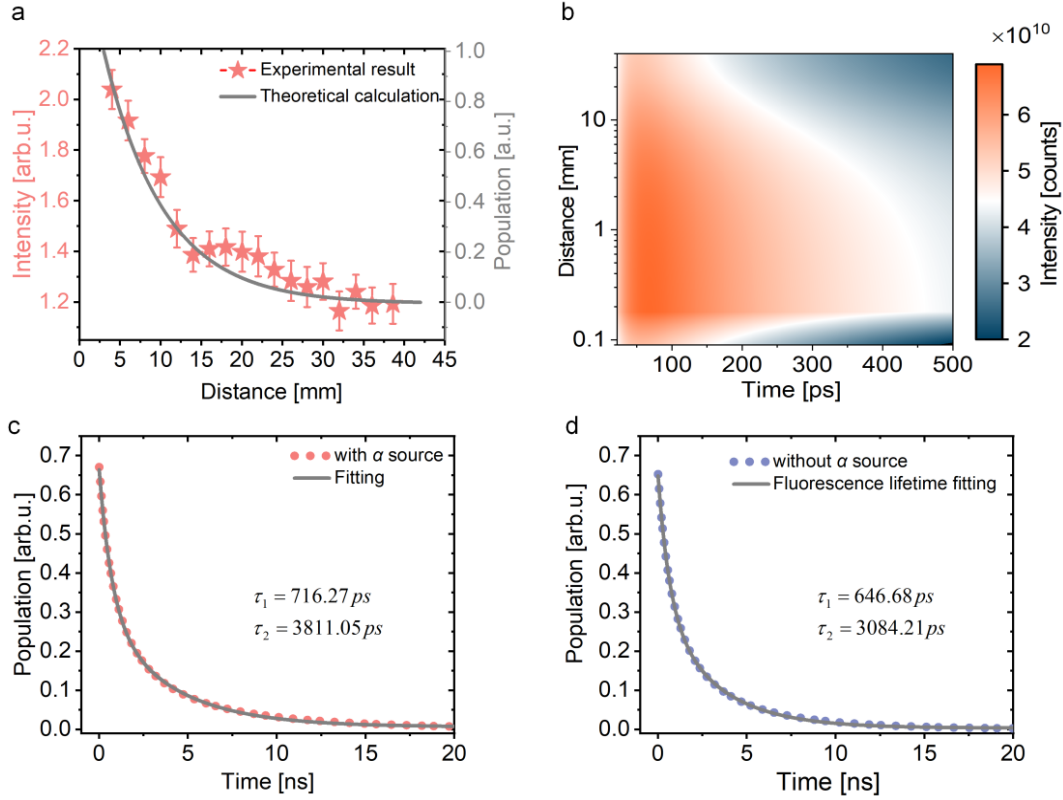

**Fig. 6. Ionizing radiation modulated femtosecond laser filamentaion dynamics.** (a) Measured (pentagonal scatter) nitrogen integral value and calculated population of $N_2(C^3\Pi_u)$ state (red curve) at different distances between α plane radiation sources and filament. (b) Calculated relaxation dynamic of the $N_2(C^3\Pi_u)$ state for different distance from filament to α source. (c), (d) Relaxation dynamic of $N_2(C^3\Pi_u)$ state for the distance at 0.4 cm: (c) α source, (d) blank plane. The simulation and double-exponential fitted results are sketched as dotted lines and solid lines, respectively.

Now we examine the formation of the $N_2(C^3\Pi_u)$ state population when considering the interaction between IR and the laser filament. The calculation results are shown in Fig. 6. The calculated population distribution of $N_2(C^3\Pi_u)$ follows the same trend as the experimentally measured nitrogen fluorescence intensity for different distances between filament and α source (Fig.6a). It can be seen from Fig. 6b that the calculated radiative lifetime of $N_2(C^3\Pi_u)$ increases as α source approaches to filament. Then we extracted and sketched the temporal evolution of $N_2(C^3\Pi_u)$ at the minimum distance between α source and filament (0.4 cm, the minimum distance in the experiments), as shown in Fig. 6c (red dots). Meanwhile, the results corresponding to blank plank (without α source) was calculated and plotted in Fig. 6d (blue dots) for comparison. Clearly, the temporal evolution of nitrogen fluorescence undergoes an ultrafast nonradiative Auger process and radiative recombination. Therefore, the double-exponential fit (grey curves) was adopted to obtain the characteristic time of the fast and slow relaxation. The slow relaxation corresponding to nitrogen fluorescence emission through radiative recombination is increased from 3 ns to 3.8 ns, which agree well with the experimental results that the extended nitrogen fluorescence lifetime of ~1 ns by $α$ radiation is observed. Such slight difference can be attributed to the limited gate width of 1 ns for our Istar-sCMOS.

**Conclusion**

In summary, we have proposed a novel filament-based ionizing radiation sensing technology (FIRST), clarified the interaction mechanism between filaments and ionizing radiation, and performed experiments based on remote detection of $α$ plate radiation sources using femtosecond laser filamentation. It has been demonstrated that when the distance between the $α$ plate radiation source and the filament is less than 1.5 cm, the electrons generated by the interaction between $α$ particles and air molecules act as seed electrons that are accelerated by the filament radiation field and collide with excited nitrogen molecules, resulting in an increase of the nitrogen fluorescence. Additionally, we found that $α$ particle ionization increased the nitrogen fluorescence lifetime of ~1 ns. These discoveries provide insight into the complex interplay between ultra-strong light fields, plasma, and energetic particle beams, opening up possibilities for the remote detection of ionizing radiation.

**Acknowledgments**

This work was supported by National Key Research and Development Program of China (2018YFB0504400).; Fundamental Research Funds for the Central Universities (63223052).

**Conflict of interest**
The authors declare no competing interests.